\documentclass[twocolumn,showpacs,preprintnumbers,amsmath,amssymb]{revtex4}
\usepackage{tabularx,graphicx}

\usepackage{color}
\usepackage{hyperref}
\hypersetup{
    colorlinks=true,
    linkcolor=blue,
    filecolor=blue,      
    urlcolor=blue,
}

\usepackage{color}

\usepackage{ulem}   

\begin{document}

\newcommand{\beq}{\begin{equation}}
\newcommand{\eeq}{\end{equation}}
\newcommand{\beqn}{\begin{eqnarray}}
\newcommand{\eeqn}{\end{eqnarray}}
\newcommand{\bmath}{\begin{subequations}}
\newcommand{\emath}{\end{subequations}}
\newcommand{\bra}[1]{\langle #1|}
\newcommand{\ket}[1]{|#1\rangle}

\title{Hole superconductivity in infinite-layer nickelates}

\author{J. E. Hirsch$^{a}$  and F. Marsiglio$^{b}$ }
\address{$^{a}$Department of Physics, University of California, San Diego,
La Jolla, CA 92093-0319\\
$^{b}$Department of Physics, University of Alberta, Edmonton,
Alberta, Canada T6G 2E1}

\begin{abstract} 
We propose that the superconductivity recently observed in Nd$_{0.8}$Sr$_{0.2}$NiO$_2$
with critical temperature in the range $9$ K to $15$ K 
results from the same charge carriers and the same mechanism that we have proposed give rise to superconductivity in 
both hole-doped and
electron-doped cuprates: pairing of hole carriers in oxygen $p\pi$ orbitals,
driven by a  correlated hopping term in the effective Hamiltonian that lowers the kinetic energy, as described by the theory of hole superconductivity. 
We predict a large increase in $T_c$ with compressive epitaxial strain.
\end{abstract}
\maketitle 

\section{introduction}

Superconductivity has been recently observed in Nd$_{0.8}$Sr$_{0.2}$NiO$_2$ \cite{nickel}.
The system has the same planes as the cuprate superconductors with Ni$^+$ instead of Cu$^{++}$ ions.
The parent compound NdNiO$_2$ is metallic at high temperatures and the resistivity turns upward at around $50$ K.
When it is doped with holes by substituting Nd$^{+++}$ by Sr$^{++}$ the resistivity continues to decrease
below $50$ K and drops to zero below the superconducting transition, which onsets around 9-15 K depending on sample 
preparation.

Key questions include:  (i) how do these compounds relate to hole-doped and 
electron-doped cuprate superconductors?
(ii) where do the doped holes go? (iii) what is the nature of the charge carriers?
(iv) what is the mechanism of superconductivity? (v) can $T_c$ be increased? In this short note we address each of these questions.

The infinite-layer phase of these materials has no apical oxygens, and is the same structure as that of infinite-layer electron-doped cuprates \cite{fournier} 
Sr$_{1-x}$Ln$_x$CuO$_2$ with  Ln = La, Nd, Sm or  Eu, which have superconducting  transition temperatures in the range 35-40 K.  
The parent compound of those electron-doped cuprates,  SrCuO$_2$, is insulating when undoped and  can only be doped with electrons, not with holes.
Instead, NdNiO$_2$, the parent compound  of these new nickelate superconductors, is metallic and (so far) can only be doped with holes.
There are also some infinite layer cuprates that appear to be hole-doped,  such as Ca$_{1-x}$Sr$_x$CuO$_2$ \cite{infhole}, but generally  hole-doped cuprate structures, 
such as La$_{1-x}$Ba$_x$CuO$_4$ contain apical oxygens \cite{holechu}.

            \begin{figure}[]
 \resizebox{6.5cm}{!}{\includegraphics[width=6cm]{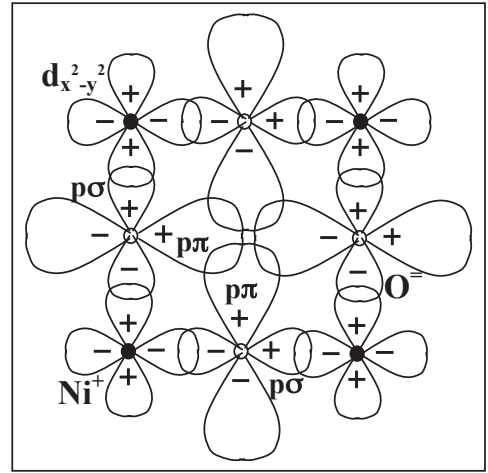}}
 \caption {Ni  d$_{x^2-y^2}$ and oxygen orbitals in the Ni-O planes. In the undoped parent compound the nominal valence is Ni$^{+}$ and
 O$^=$ and there is one hole in the filled Ni $d^{10}$  orbital. The O-$p\pi$ orbitals point in a direction perpendicular to the
 Ni-O bonds, and the $p\sigma$ orbitals in a direction that is parallel. We propose that doped holes reside in a band resulting principally from overlapping 
 O-$p\pi$ orbitals, the same as for  for both hole- and electron-doped cuprates.
 }
 \label{figure1}
 \end{figure} 

It is easy to understand these doping characteristics. LaCuO$_4$ can be doped with holes because the presence of 
O$^{--}$ apical oxygens lowers the electrostatic potential at the Cu$^{++}$ site. Another way to lower the electrostatic potential
at the cation site is to replace Cu$^{++}$ by Ni$^{+}$, without apical oxygens. For that reason, these new nickelates \cite{nickel} can be doped
with holes. Instead, with Cu$^{++}$ and no apical oxygens
the electrostatic potential is higher so it is easier to dope with electrons rather than with holes.  

Notwithstanding the type of dopant charges,   we have proposed that superconductivity in hole-doped \cite{hdoped} and electron-doped \cite{edoped1}  cuprates arises from
{\it the same carriers} through the same mechanism \cite{edoped}. 
 The fact that the same type of carriers appear to be responsible for the superconductivity of
hole-doped and electron-doped cuprates has in our view  been firmly established by extensive experimental evidence
  \cite{wang,crusellas,greene1,greene2,greene4,greene5,greven}.
The newly discovered superconducting nickelates share with the hole-doped cuprates the fact
that they are hole-doped, and share with electron-doped cuprates the fact that the structure has no apical oxygens.
It  would not be very  surprising if  the same carriers and the same mechanism account for the superconductivity of
these new materials also.

 Figure 1 shows the relevant orbitals in the NiO (or CuO) planes. The consensus for the cuprates is that both under
 hole or electron doping the doped carriers go into the Cu-$d_{x^2-y^2}$ O-${p \sigma}$ band, and give rise to superconductivity.
 Instead, we have  proposed that it is always holes in O-${p \pi}$ orbitals that give rise to superconductivity.
 For electron-doped cuprates, electrons go into the Cu-$d_{x^2-y^2}$ O-${p \sigma}$ band and  induce holes into the O-${p \pi}$ 
 orbitals \cite{edoped}.
 Instead, for hole-doped cuprates and nickelates, the doped holes go directly into the O-${p \pi}$ band.
 We have argued that for both hole- and electron-doped cuprates orbital relaxation
 of the highly negatively charged oxygen anion lifts the O-${p \pi}$ orbitals to the Fermi level \cite{cuo}, contrary to what
 band structure calculations predict. Clearly the same argument applies here.

Figure 2 shows the energy level structure we envision for cuprates and nickelates, in a hole representation. 
Both in  the hole-doped cuprates (left panel) and in the hole-doped nickelates (right panel) the
added hole will go into the oxygen $p \pi$ level because adding it to the
cation would cost a  high Hubbard $U$. 
In the electron-doped cuprates (center panel), removing a hole (adding an electron) causes the other hole to `fall' into
the O-$p\pi$ orbital.  
In the hole-doped infinite layer nickelates removing a hole (adding an electron)
costs more energy than in the electron-doped infinite layer cuprates, so
it is not likely that  these new materials can be doped with electrons.
            \begin{figure}[]
 \resizebox{8.5cm}{!}{\includegraphics[width=6cm]{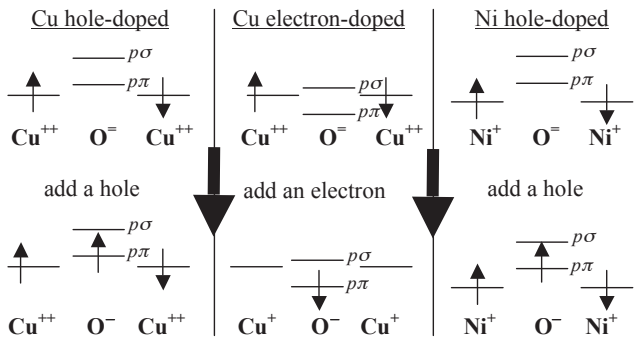}}
 \caption {Illustration of how oxygen $p\pi$ hole carriers are created in hole-doped cuprates,  electron-doped
 cuprates, and  hole-doped nickelates,   in the hole representation. Arrows on the energy levels denote holes. The difference in the relative locations of the 
 O$^=$ and Cu$^{++}$ orbitals for  hole-doped and electron-doped cases arises due to their different
 crystal structures, T  versus  T$^\prime$ or infinite layer.
 For the nickelates, the Ni$^+$ level is lower relative to the oxygen levels compared to the electron-doped cuprates due to the lower atomic number.
 }
 \label{figure1}
 \end{figure} 
\begin{figure}[]
 \resizebox{7.5cm}{!}{\includegraphics[width=6cm]{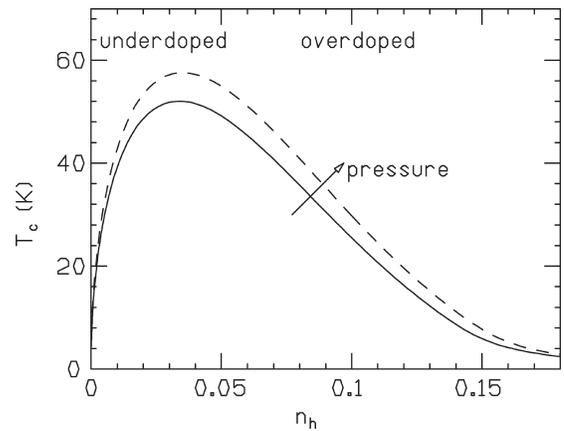}}
 \caption {  $T_c$ versus hole concentration $n_h$, the
number of  holes per oxygen atom. This behavior is generic for
this model. Values for the bandwidth, correlated hopping parameter,
on-site and nearest neighbor repulsion used are $D$ = 5 eV,
$t$ = 0.3725~eV, $U$ = 5 eV, and $V$ = 0, respectively. The dashed line
indicates the behavior expected under application of  pressure in the plane, with the hopping parameter
 $t$ increased to 0.375~eV. }
 \label{figure1}
 \end{figure} 
 Experiments have so far been reported for one hole concentration.
 Figure 3 shows the predicted dependence of $T_c$ on hole concentration for a typical set of parameters in our model.
 Within this model  $T_c$ increases rapidly when the distance between atoms in the plane decreases,
 as we showed in Ref.~\cite{hdoped}. Therefore we predict a large enhancement of $T_c$ under compressive epitaxial strain.
 The dashed line in Fig. 3 shows the expected behavior of $T_c$ versus hole concentration under such
 strain, with an increase in $T_c$ at {\it all hole concentrations}.

 We expect the Hall coefficient in this material to be positive at low temperatures, reflecting conduction of holes
 in a single nearly full band. Figure 4 shows the measured Hall coefficient \cite{nickel}, which turns positive below 50 K.
At higher temperature the negative Hall coefficient presumably results from contribution to conduction from
both the Ni-O $p\sigma$ band and  the O-$p\pi$ band, with higher mobility for
the carriers in the Ni-O band.  Note that in the undoped case 
the conduction is metallic above 50K (Fig. 3b in Ref. \cite{nickel}), presumably due to 
the smaller on-site repulsion $U$ compared to the cuprates, and the Hall coefficient is negative as seen in Fig. 4.
 
Within our model, superconductivity is caused  by a correlated hopping term in the effective Hamiltonian describing conduction
of oxygen $p\pi$ holes in a nearly full band \cite{hdoped}, which arises due to the contraction and expansion of the 
oxygen orbitals depending on their charge content \cite{dynhub}. Superconductivity is driven by lowering of quantum kinetic energy, and neither phonons nor
spin fluctuations play a significant role.  It requires conduction by holes. 
We have proposed that this is also the mechanism responsible for
superconductivity in iron pnictides \cite{feas}, magnesium diboride \cite{mgb2,mgb22},
H$_2$S \cite{h2s} and all other superconducting materials \cite{matmech}. Correlated hopping has also recently been proposed to be 
responsible for superconductivity in twisted bilayer graphene \cite{guinea} and for superfluidity in systems of ultracold atoms in
optical lattices \cite{atoms,atoms2}. 
 \begin{figure}[]
 \resizebox{7.5cm}{!}{\includegraphics[width=6cm]{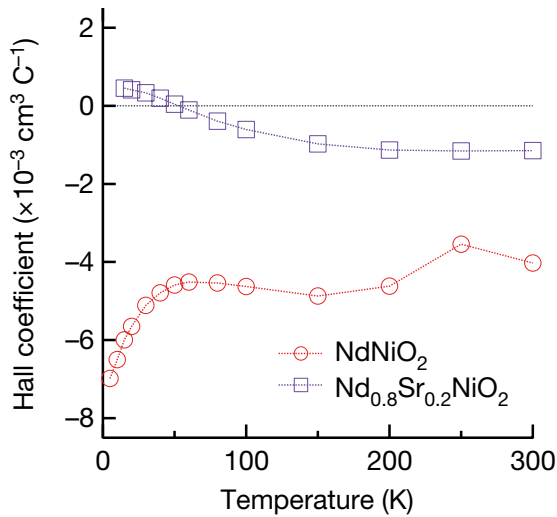}}
 \caption {Measured Hall coefficient in undoped and doped superconducting nickelates, from Ref. \cite{nickel}.
 }
 \label{figure1}
 \end{figure} 
 
 As pointed out in Ref.~\cite{nickel} and also emphasized  by Sawatzky \cite{sawatzky}, the discovery of superconductivity in these nickelates 
 poses a challenge to proposed explanations of cuprate superconductivity that rely on magnetism, Zhang-Rice singlets, 
 Mott physics, RVB, large Hubbard $U$, spin fluctuations,  etc. Even though in elemental form Ni is magnetic,
 there is so far no evidence for magnetism in the newly discovered nickelates, and the parent undoped compound
 is metallic rather than insulating.  In some sense these compounds appear closer to the class of superconducting bismuthates \cite{sleight},
 for which both conventional \cite{bismuthconv,bismuthconv2} and `negative U' \cite{geballenegu} mechanisms have been proposed. Of course in other respects, e.g. structurally, the nickelates
are much closer to the cuprates. Botana and Norman \cite{norman} have argued that the electronic structure of nickelates is similar to that
 of cuprates and suggested a common mechanism. Others would surely  argue that so far at least  there is no evidence that the
 nickelates are anything other than conventional BCS-electron-phonon superconductors.
 
 In summary, it is clear that the discovery of the infinite-layer nickelates \cite{nickel} with a significant superconducting critical temperature
 adds an important new member to a growing number  of superconducting materials classes \cite{spissue}. 
 Commonalities and differences between the different classes  \cite{spissueintro} should help to significantly narrow down the range of plausible theories.
 In our view, most compelling is the reported Hall coefficient
 reproduced in our Fig.~4, which indicates the importance of hole carriers also in this new class. Measurements of the superconducting $T_c$ 
 and Hall coefficient at other hole
 concentrations will help to support our model, as will experiments reporting on the pressure dependence of $T_c$
 and on tunneling asymmetry \cite{hdoped,edoped}.
 We don't see any reason $not$ to expect significantly higher transition temperatures in this new class of 
 oxide superconductors, particularly if the in-plane lattice constant can be reduced and the carrier concentration optimized.

\begin{acknowledgments}

This work was supported in part by the Natural Sciences and Engineering
Research Council of Canada (NSERC).  

\end{acknowledgments}

 \end{document}